\documentclass{jpsj3}
\usepackage{txfonts}
\usepackage{ascmac}
\title{Interference Pattern Formation between Bound Solitons and Radiation in Momentum Space: Possible Detection of Radiation from Bound Solitons with Bose-Einstein Condensate of Neutral Atoms}

\author{
Hironobu F{\sc ujishima},$^{1}$\thanks{E-mail address:
fujishima.hironobu@canon.co.jp} Masahiko O{\sc kumura},$^{2,3,4,5}$\thanks{E-mail address:
okumura.masahiko@jaea.go.jp} Makoto M{\sc ine},$^6$\thanks{E-mail address:
mine@waseda.jp} and Tetsu Y{\sc ajima}$^7$\thanks{E-mail address:
yajimat@is.utsunomiya-u.ac.jp}
}
\inst{
$^1$Optics R\&D Center, CANON INC., 23-10 Kiyohara-Kogyodanchi,
Utsunomiya 323-3298, Japan \\
$^2$CCSE, Japan Atomic Energy Agency, 5-1-5 Kashiwanoha, Kashiwa, Chiba 277-8587, Japan\\ 
$^3$CREST (JST), Kawaguchi, Saitama 332-0012, Japan \\
$^4$Computational Condensed Matter Physics Laboratory, RIKEN, Wako, Saitama 351-0198, Japan\\
$^5$Computational Materials Science Research Team, RIKEN AICS, Kobe, Hyogo 650-0047, Japan\\   
$^6$Waseda University Honjo Senior High School, Kurisaki 239-3, Honjo, Saitama 367-0032, Japan \\
$^7$Department of Information Systems Science, Graduate School of
Engineering, Utsunomiya University, Yoto 7-1-2, Utsunomiya 321-8585,
Japan 
}
\abst{We propose an indirect method for observing radiation from an incomplete soliton with a sufficiently large amplitude. We show that the radiation causes a notched structure on the envelope of the wave packet in the momentum space. The origin of this structure is the interference between the main body of oscillating solitons and the small radiation in the momentum space. We numerically integrate the nonlinear Schr\"odinger equation and perform Fourier transformation to confirm that the predicted structure really appears. We also show a simple model which reproduces the qualitative result. The experimental detection of the notched structure with the Bose-Einstein condensation of neutral atoms is discussed and suitable parameters for this detection experiment are shown. 
}
\kword{one-dimensional nonlinear Schr\"odinger equation, one-dimensional Gross-Pitaevskii equation, radiation, interference, numerical analysis,
Bose-Einstein condensation} 
\begin{document}
\maketitle

\section{Introduction}
The one-dimensional nonlinear Schr\"odinger equation (NLSE),
\begin{equation}
\mathrm{i}\psi_{t}=-\frac{1}{2}\psi_{xx}-|\psi|^2\psi\label{NLSE},
\end{equation}
is a well-known soliton equation which appears in various fields of physics\cite{Karpman} and has many applications, such as in optical fiber communication\cite{Hasegawa,Agrawal}. Another important example of physical systems for the NLSE is seen in the Bose-Einstein condensation (BEC) of neutral atoms,\cite{Dalfovo,Pitaevskii} where the macroscopic wave function of condensate atoms appears as the order
parameter accompanied by the spontaneous breakdown of the $U(1)$
gauge symmetry\cite{Okumura}. In this case, the NLSE is regarded as the mean-field approximation of the Heisenberg equation for field
operators and describes the time evolution of this macroscopic wave
function with good accuracy\cite{Dalfovo,Pitaevskii}. Up to this day, both bright and dark soliton-like objects have already been created with BEC\cite{Khaykovich,Burger} (though the dark solitons are outside the scope of this study.). 
 
 On the other hand, it is clear that we cannot realize a rigorous soliton initial condition in any real experiment. However carefully prepared, any realized physical object must be accompanied by deviation from ideal solitons. Even in such cases, it is widely known that the NLSE has analytical integrability; exact solutions under an arbitrary initial condition are made available by the inverse scattering transformation (IST) method\cite{Gardiner,Zakharov}. In particular, to analyze soliton phenomena, we can exploit Hirota's direct method\cite{Hirota}. The IST method gives us not only the solutions of the NLSE but also a clear outlook on their classifications.
It is based on an auxiliary linear
eigenvalue problem, 
\begin{equation}
\begin{pmatrix}
\Psi_{1x}\\
\Psi_{2x}
\end{pmatrix}=
\begin{pmatrix}
-\mathrm{i}\zeta &\mathrm{i}\psi^{\ast} \\
\mathrm{i}\psi &\mathrm{i}\zeta
\end{pmatrix}
\begin{pmatrix}
\Psi_{1}\\
\Psi_{2}
\end{pmatrix}\label{eigen-1},
\end{equation}
where $\zeta$ is the eigenvalue which is independent of time, $\psi$
is the solution of the NLSE, and the asterisk means its complex
conjugate. This formulation takes the form of a potential scattering
problem for the auxiliary fields $\Psi_1$ and $\Psi_2$, where $\psi$ works as a
potential. The eigenvalue spectrum consists of discrete and continuous
parts, where the former generates soliton solutions and the latter
corresponds to small ripples called ``radiation''. For a sech-type initial condition,
\begin{equation}
\psi(x,0)=\mathit{M}\mathrm{sech}(x)\label{sech},
\end{equation}
the eigenvalue consists of a discrete part only, provided that $\mathit{M}$ equals a positive integer. In
this case, the whole initial value problem is solved analytically and
results in an $\mathit{M}$-soliton bound state\cite{Satsuma}.

The time evolution of the auxiliary field is defined by another linear equation,
\begin{equation}
\begin{pmatrix}
\Psi_{1t}\\
\Psi_{2t}
\end{pmatrix}=
\begin{pmatrix}
\mathrm{i}\zeta^2-\frac{1}{2}\mathrm{i}|\psi|^2 &\frac{1}{2}\psi^{\ast}_{x}-\mathrm{i}\zeta\psi^{\ast} \\
-\frac{1}{2}\psi_{x}-\mathrm{i}\zeta\psi &-\mathrm{i}\zeta^2+\frac{1}{2}\mathrm{i}|\psi|^2
\end{pmatrix}
\begin{pmatrix}
\Psi_{1}\\
\Psi_{2}
\end{pmatrix}\label{eigen-2}.
\end{equation}
However, the time evolution of the wave packet from initial conditions except eq. (\ref{sech}) is relatively unclear since the analytic treatment of the continuous spectrum is difficult. All we know is the asymptotic behavior. According to the analytic results obtained by Satsuma and Yajima\cite{Satsuma}, a solitary pulse is very robust and, unless the spectrum of the eigenvalue problem (\ref{eigen-1}) contains a discrete eigenvalue, the wave packet finally splits into the soliton and radiation parts. The norm of the soliton part remains $O(1)$. When the amplitude of the wave packet is sufficientry large, the bound soliton can be formed asymptotically and has already been observed in real experiments with optical fibers\cite{Mollenauer}. In contrast, the amplitude of the radiation part decays as $O(t^{-\frac{1}{2}})$ and expands rapidly from the main body of the soliton part.

In this way, the amplitude of radiation is generally very small, which makes it difficult to observe radiation in real experiments. In fact, whatever the research field is, no clear observation of radiation has been reported. The study of radiation has been difficult both theoretically and experimentally, and the problems on the dynamics of the NLSE, starting from general initial conditions, have not been solved. 

In this paper, we propose an idea toward the experimental observation of radiation through pattern formation in momentum space. We exploit the function of order formation originating from the nonlinearity of the NLSE.    

This paper is organized as follows. In the next section, our theoretical idea is explained and the results of numerical simulations are shown. In \S 3, the feasibility of a real experiment is discussed. Section 4 is devoted to discussion and summary. 

\section{Theory and Numerical Results}
\begin{figure}[h]
\begin{center}
\includegraphics[width=80mm]{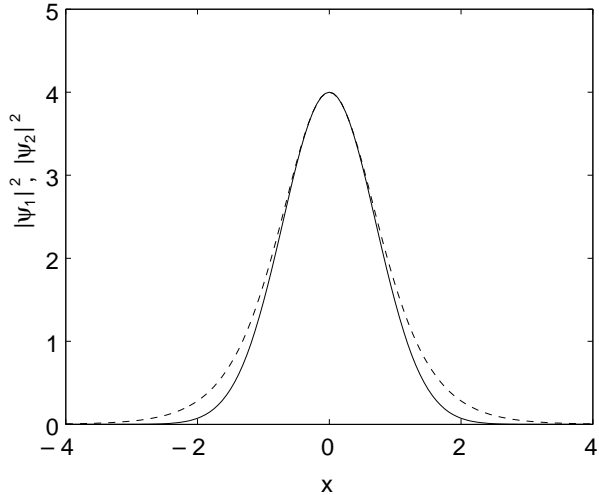}
\end{center}
\caption{Profiles of the initial wave packets. The dashed curve is for intial condition (\ref{init1}) and the solid curve is for initial condition (\ref{init2}).} \label{f1}
\end{figure}
\begin{figure}[h]
\begin{center}
\includegraphics[width=80mm]{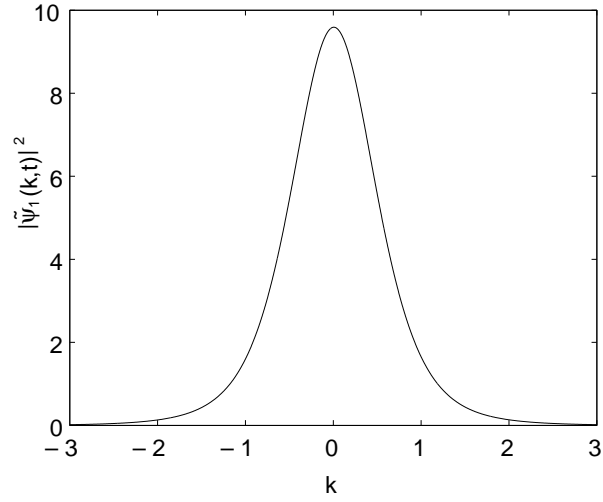}
\end{center}
\caption{Momentum space profile of the wave packet $|\tilde{\psi}_{1}(k,t)|^2$ starting from initial condition (\ref{init1}) at $t=14$.} \label{f2}
\end{figure}
\begin{figure}
\begin{center}
\includegraphics[width=80mm]{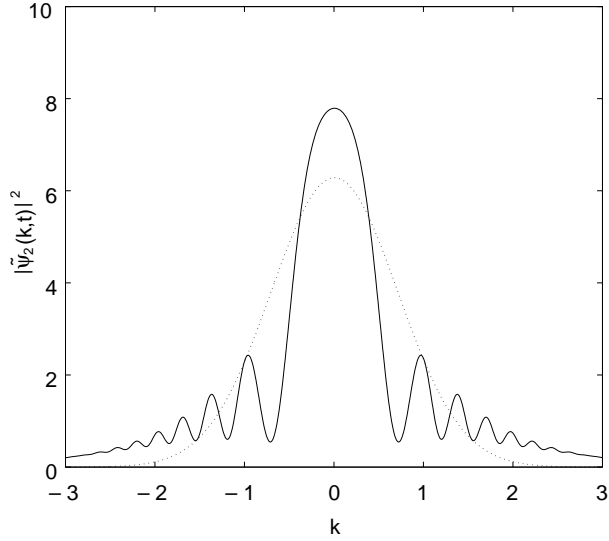}
\end{center}
\caption{Momentum space profile of the wave packet $|\tilde{\psi}_{2}(k,t)|^2$ starting from initial condition (\ref{init2}) at $t=14$. The dashed curve is for the initial profile.} 
\label{f3}
\end{figure}
\begin{figure}
\begin{center}
\includegraphics[width=80mm]{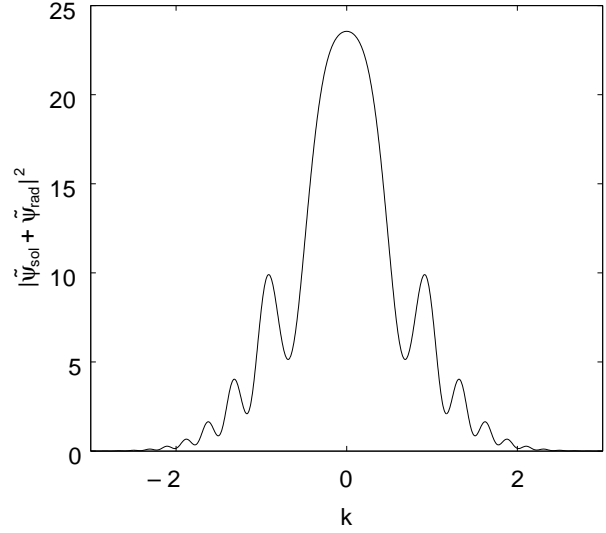}
\end{center}
\caption{Momentum space profile of the wave packet $|\tilde{\psi}_{\rm sol}+\tilde{\psi}_{\rm rad}|^2$, where $\tilde{f}(k)$ is taken to be $\mathrm{e}^{-\frac{1}{2}k^2}$. This snapshot was taken at $t=14$.} 
\label{f4}
\end{figure}
\begin{figure}
\begin{center}
\includegraphics[width=80mm]{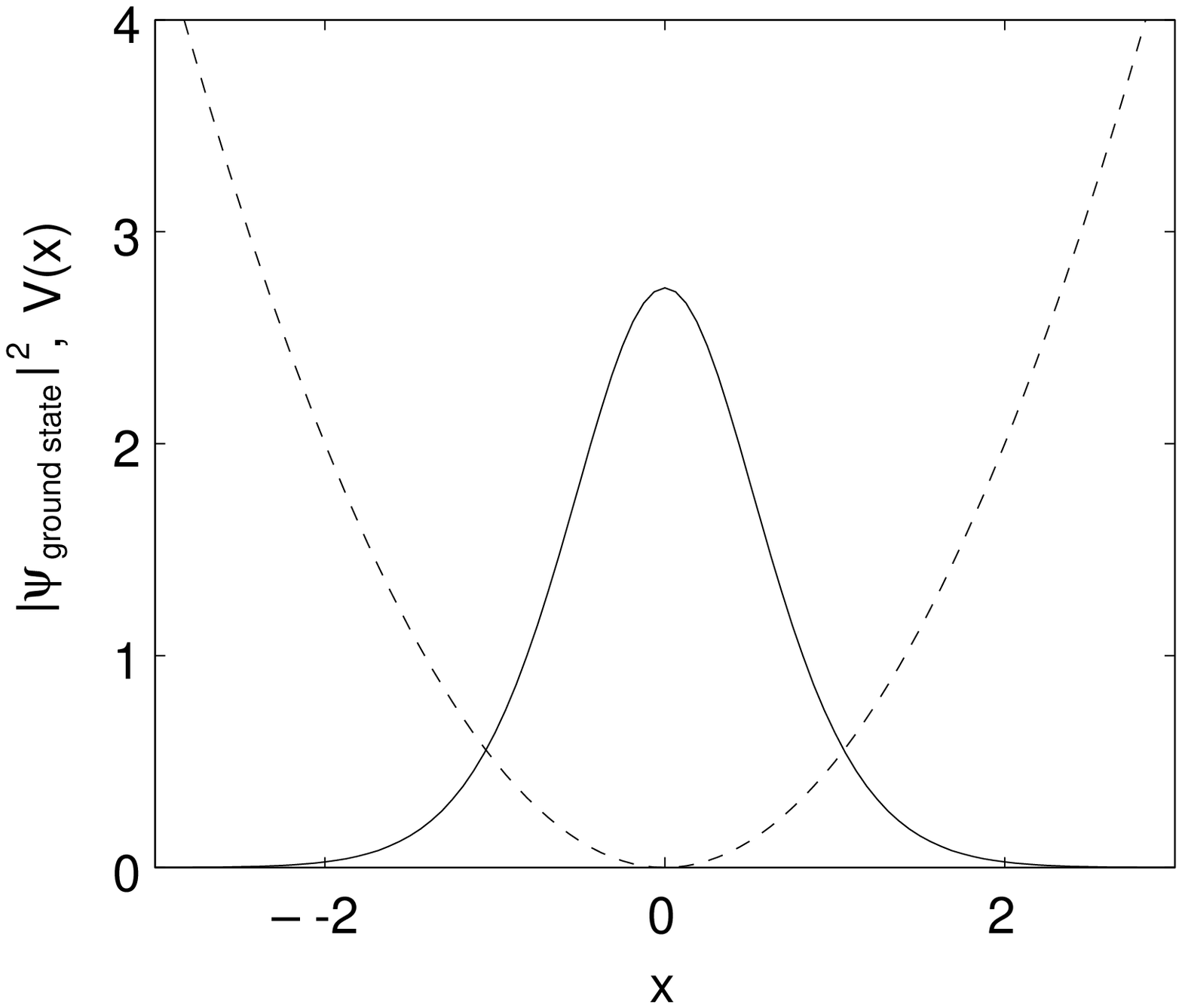}
\end{center}
\caption{Profile of the ground state wave packet in the external potential $V_{\mathrm{ext}}(x)=\frac{1}{2}x^2$. The real line is for the profile of the wave packet and the dashed line is for the external potential $V_{\mathrm{ext}}(x)$.} 
\label{f5}
\end{figure}
\begin{figure}
\begin{center}
\includegraphics[width=80mm]{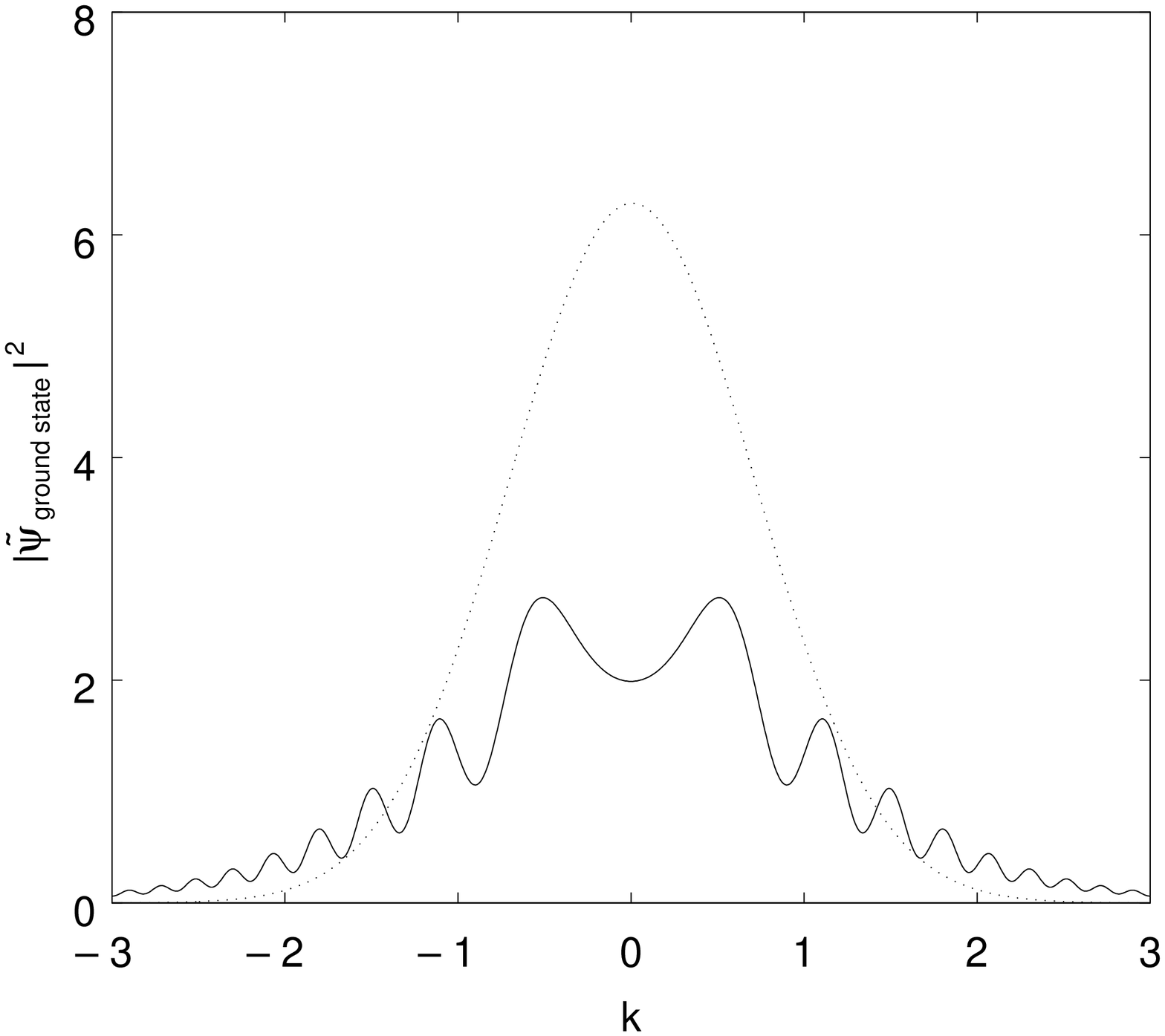}
\end{center}
\caption{Momentum space profile of the wave packet starting from the ground state one in the external potential $V_{\mathrm{ext}}(x)=\frac{1}{2}x^2$. This snapshot was taken at $t=14$. The dashed curve is for the initial profile.} 
\label{f6}
\end{figure}
\begin{figure}
\begin{center}
\includegraphics[width=80mm]{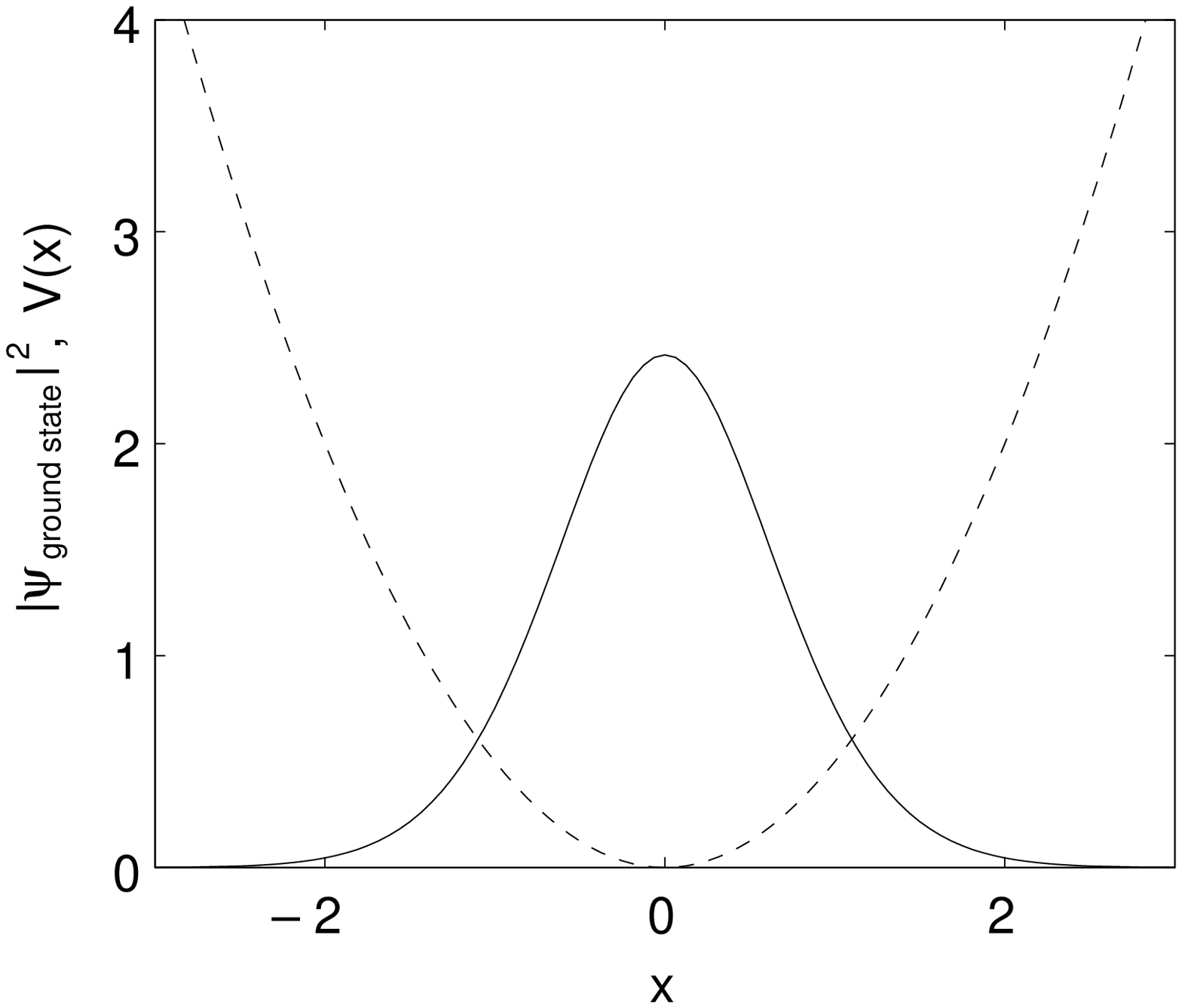}
\end{center}
\caption{Profile of the ground-state wave packet in the external potential $V_{\mathrm{ext}}(x)=\frac{1}{2}x^2$. The real line is for the profile of the wave packet for $g'_{1D}=0.7$, and the dashed line is for the external potential $V_{\mathrm{ext}}(x)$.} 
\label{f7}
\end{figure}
\begin{figure}
\begin{center}
\includegraphics[width=80mm]{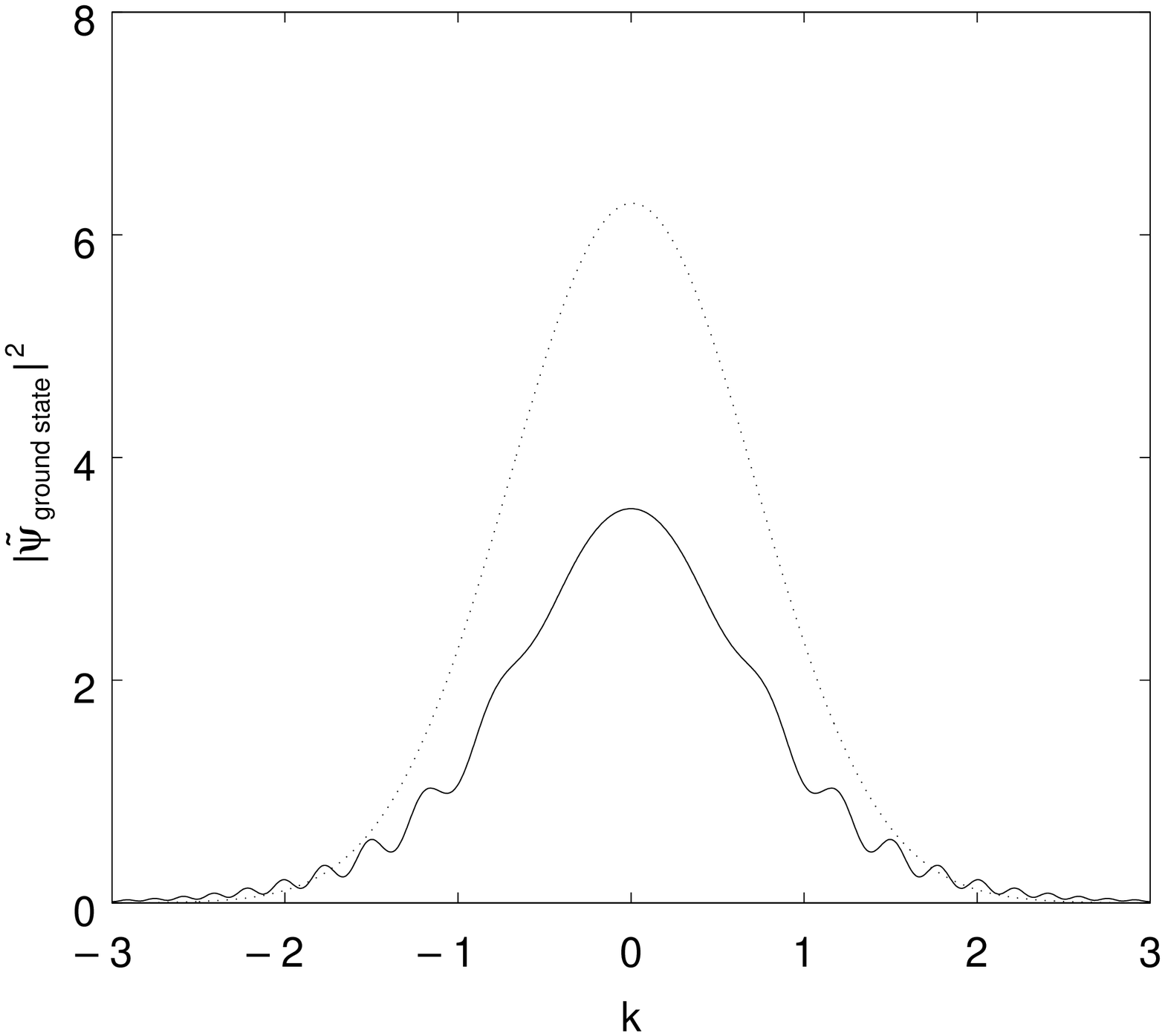}
\end{center}
\caption{Momentum space profile of the wave packet starting from the ground-state one in the external potential $V_{\mathrm{ext}}(x)=\frac{1}{2}x^2$. This snapshot was taken at $t=14$ with $g'_{1D}=0.7$. The dashed curve is for the initial profile.} 
\label{f8}
\end{figure}
\begin{figure}
\begin{center}
\includegraphics[width=80mm]{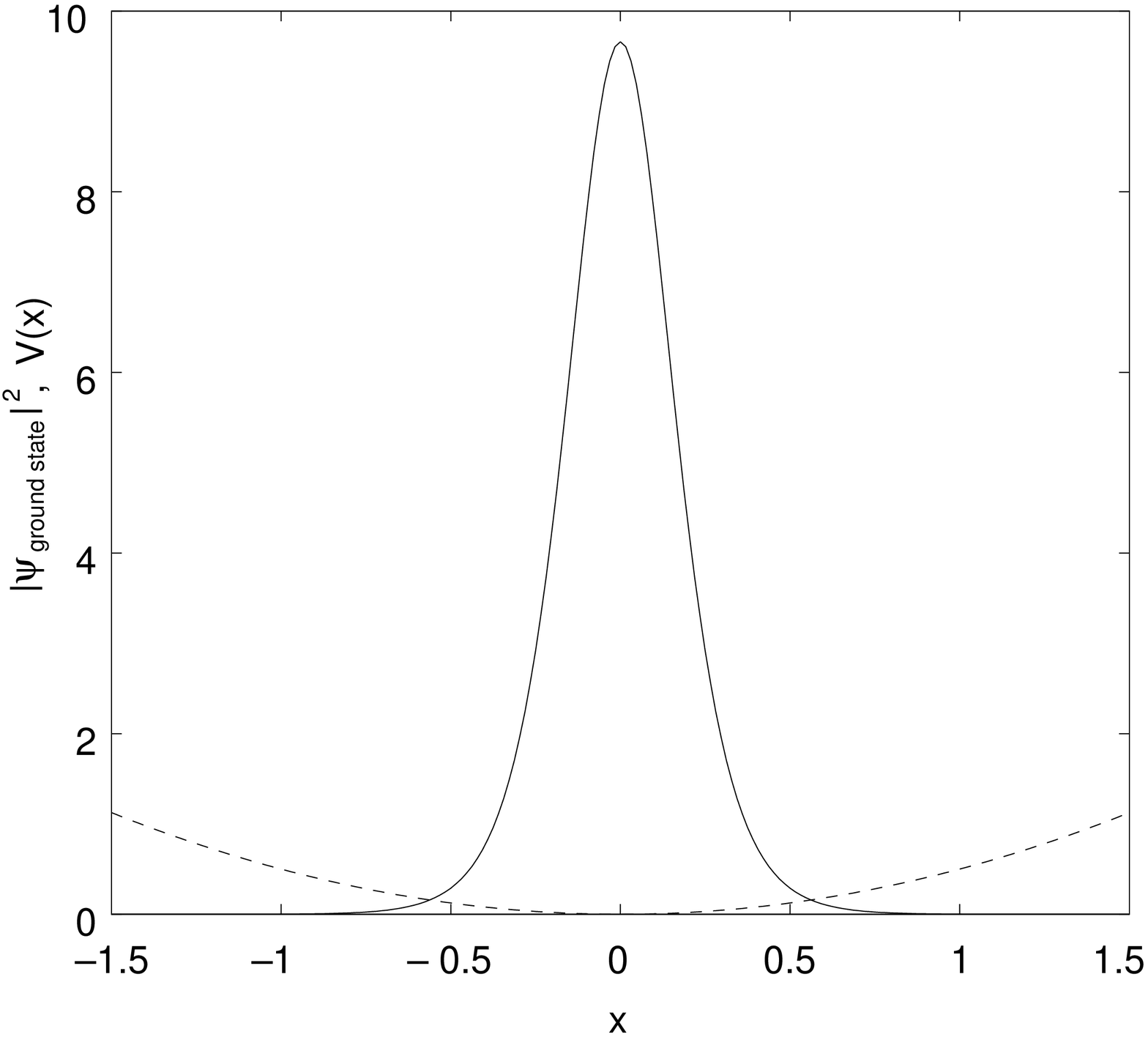}
\end{center}
\caption{Profile of the ground-state wave packet in the external potential $V_{\mathrm{ext}}(x)=\frac{1}{2}x^2$. The solid line is for the profile of the wave packet for  $g'_{1D}=5$, and the dashed line is for the external potential $V_{\mathrm{ext}}(x)$.}  
\label{f9}
\end{figure}
\begin{figure}
\begin{center}
\includegraphics[width=80mm]{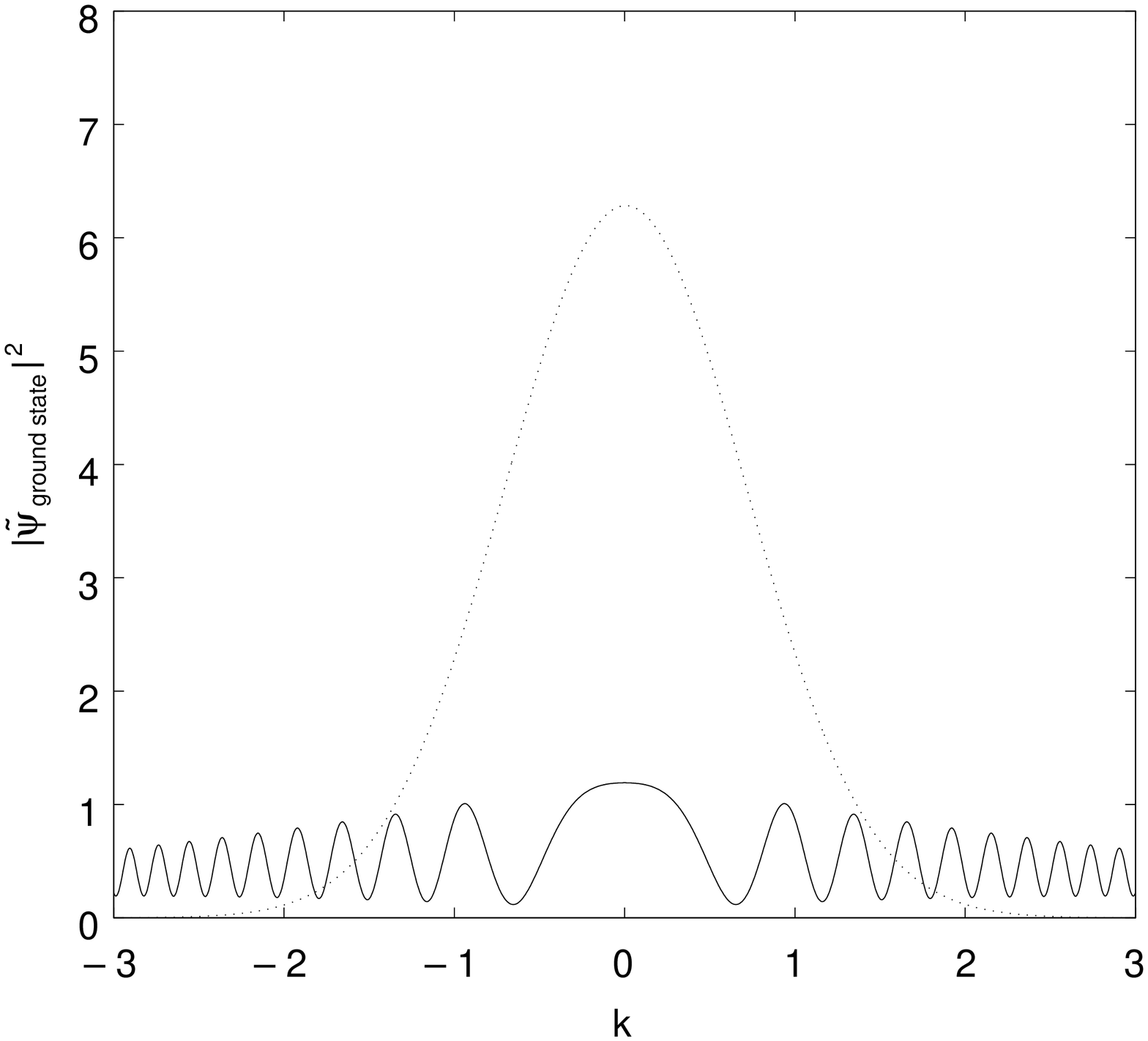}
\end{center}
\caption{Momentum space profile of the wave packet starting from the ground-state one in the external potential $V_{\mathrm{ext}}(x)=\frac{1}{2}x^2$. This snapshot was taken at $t=14$ with $g'_{1D}=5$. The dashed curve is for the initial profile.} 
\label{f10}
\end{figure}
In this section, we briefly summarize the mathematical descriptions of
the system to be considered. Throughout this paper, we consider the NLSE (\ref{NLSE}) under initial conditions, the eigenvalue spectrum of which contains two or more discrete eigenvalues. By virtue of the Galilei transformation, we can restrict ourselves to the case where the center of the wave packet remains at $x=0$. Such a situation is achieved as long as the initial conditions are real valued and symmetric functions of $x$. Under such conditions, the relative velocities of the contained solitons vanish and these solitons form bound states.

We consider two similar initial conditions which meet the requirement above. The first initial condition is the sech-type (\ref{sech}) one with $\mathit{M}=2$,
\begin{equation}
\psi_{1}(x,0)=2\mathrm{sech}(x)\label{init1}.
\end{equation}
Starting from $\psi_{1}(x,0)$ results in a pure two-soliton bound state, and the solution of which reads,
\begin{equation}
\psi_{1}(x,t)=4\mathrm{exp}(-\frac{\mathrm{i}t}{2})\frac{\mathrm{cosh}(3x)+3\mathrm{exp}(-4\mathrm{i}t)\mathrm{cosh}(x)}{\mathrm{cosh}(4x)+4\mathrm{cosh}(2x)+3\mathrm{cos}(4t)}.\label{2sol}
\end{equation}
The envelope of this solution pulsates with the frequency $\frac{\pi}{2}$ and no radiation is emitted.

The second initial conditon is of the Gaussian type with the same amplitude and a similar width,
\begin{equation}
\psi_{2}(x,0)=2\mathrm{exp}(-\frac{1}{2}x^2)\label{init2}.
\end{equation}
This form of initial condition not only deviates from the complete soliton but also becomes one of the accessible forms, especially in BEC experiments, and its time evolution was investigated in detail in our previous work\cite{Fujishima}. By the Feshbach resonance technique\cite{Courteille}, we can erase the effective interatomic interaction, and the atoms trapped in a quadratic potential trap take the ground state to form the Gaussian profile.
We show $|\psi_{1}(x,0)|^2$ and $|\psi_{2}(x,0)|^2$ in Fig.~\ref{f1}. They show similar curves; in fact, both lead to pulsars. However, the wave packet starting from initial condition (\ref{init2}) does not have a simple analytical expression and keeps emitting a very small amount of radiation which has not been observed experimentally.

We propose a method of distinguishing these two initial conditions.

For the above purpose, we observe that there appears a distinguishable feature on the envelope of a wave packet which emits radiation. That is, a ``notched structure" is formed on the envelope in the momentum space. The creation of the notched structure is explained as follows.

When we consider an initial condition with a sufficiently large amplitude that deviates from that of the complete soliton, the total wave function can be expressed as the sum of a soliton part $\psi_{\rm sol}$ and a radiation part $\psi_{\rm rad}$, i.e., $\psi_{\rm sol}+\psi_{\rm rad}$. We write these two constituent wave functions as $\psi_{\rm sol}=g(x)e^{i\theta(x)}$ and $\psi_{\rm rad}=h(x)e^{i\theta(x)}$, where $g(x)$ and $h(x)$ are absolute values of $\psi_{\rm sol}$ and $\psi_{\rm rad}$, respectively. They have a definite phase $\theta(x)$ in common; thus, they never interfere with each other in the real space. However, when we perform the Fourier transformation on these wave functions, the difference between $g(x)$ and $h(x)$ induces two different phases for $\tilde{\psi}_{\rm sol}$ and $\tilde{\psi}_{\rm rad}$, where the tilde means a Fourier component. Now, they can interfere with each other in the momentum space. This interference pattern in the momentum space is nothing else but our notched structure mentioned above and this is an outline of our strategy. Below, we show a brief calculation of the expression of the interference pattern on the bases of a rough but concise model.

Since the pulsating soliton part localizes around the origin and oscillates with monochromatic frequency $\omega=\frac{\pi}{2}$, it can be approximated as
\begin{equation}
\psi_{\rm sol}(x,t)=\mathrm{e}^{-\frac{1}{2}x^2}(3+\mathrm{cos}(4t)).\label{app}
\end{equation}
We note that this approximation holds only around the origin; in fact, when $\psi(x,t)$ shrinks, the rest part of the main body of solitons extends toward the left and right to conserve the whole norm. However, this rather rough approximation does not spoil the qualitative argument because the oscillating amplitude of the side lobe is much smaller than that of the main body of solitons near the origin. 

On the other hand, we can neglect the nonlinear interaction term of the radiation part because its amplitude is very small.
Therefore, the radiation part is well described by the linear Schr\"odinger equation.
Such a wave motion with zero group velocity is expressed as  
\begin{equation}
\psi_{\rm rad}(x,t)=\int\tilde{f}(k)\mathrm{e}^{-\frac{\mathrm{i}}{2}k^{2}t}\mathrm{e}^{\mathrm{i}kx}dk,\label{LAD}
\end{equation}
where $\tilde{f}(k)$ is a real and symmetric function of $k$.This expression for the radiation part is no more than an approximation. In fact, there exists an interaction between the radiation part and the soliton part, particularly near the origin where the amplitudes of these parts are large. In such an area, the approximation is poor. However, such an area is relatively very narrow and limited to around the origin. To observe the notched structure, we perform the Fourier transformation on the wave function in the real space. By virtue of Fourier transformation, which is an integral over an infinitely long interval, the local error is smeared or averaged out in the momentum space and brings about no crucial alternation but a small offset to our original idea. In almost all othe areas but the origin, we have confirmed numerically that our approximation is valid.

We shall consider the interfered profile of these soliton and radiation parts in momentum space,
\begin{equation}
|\tilde{\psi}_{\rm sol}+\tilde{\psi}_{\rm rad}|^2.\label{super}
\end{equation}
For the validity of superposing these approximated functions, see the appendix. Obviously, we find $\tilde{\psi}_{\rm sol}(k,t)=\mathrm{e}^{-\frac{1}{2}k^2}(3+\mathrm{cos}(4t))$ and $\tilde{\psi}_{\rm rad}(k,t)=\tilde{f}(k)\mathrm{e}^{-\frac{\mathrm{i}}{2}k^{2}t}$.
Therefore,
\begin{equation}
|\tilde{\psi}_{\rm sol}+\tilde{\psi}_{\rm rad}|^2=(3+\mathrm{cos}(4t))^{2}\mathrm{e}^{-k^2}+\tilde{f}^{2}(k)+2(3+\mathrm{cos}(4t))\tilde{f}(k)\mathrm{e}^{-\frac{1}{2}k^2}\mathrm{cos}(\frac{1}{2}k^2 t).\label{result}
\end{equation}
This interference pattern is observed as the notched structure on the surface profile of the wave packet in momentum space. The larger $k$ makes the oscillation period shorter and the structure is sent out of the center of the profile. This spatial and temporal order is quite outstanding and it makes it possible for us to observe radiation indirectly. 

We show the results of numerical simulation based on the model explained above.

In Figs.~\ref{f2} and \ref{f3}, the snapshots of $|\tilde{\psi}_{1}(k,t)|^2$ and $|\tilde{\psi}_{2}(k,t)|^2$ are depicted at $t=14$. Although the surface of $|\tilde{\psi}_{1}(k,t)|^2$ is smooth,
the notched structure is fully developed on the surface of $|\tilde{\psi}_{2}(k,t)|^2$. For reference, Fig.~\ref{f4} shows $|\tilde{\psi}_{\rm sol}+\tilde{\psi}_{\rm rad}|^2$ at $t=14$, where $\tilde{f}(k)$ is taken to be $\mathrm{e}^{-\frac{1}{2}k^2}$. It is seen that our model qualitatively well describes the result in Fig.~\ref{f3} even though the norm becomes much larger because our simple model does not preserve the norm.

Next, we consider a more general case. In real experiments, one usually trap atoms with an external potential $V_{\mathrm{ext}}(x)$, which can be approximated by a quadratic function. During the trapping time, the macroscopic wave function of atoms goes to the ground-state. The ground-state wave function in the potential corresponds to the initial condition. After this process in the trapping potential, we switch off the trap and release the atoms to observe the interference pattern. Figure~\ref{f5} shows the profile of the ground-state wave packet in the external potential $V_{\mathrm{ext}}(x)=\frac{1}{2}x^2$. The real line is for the profile of the wave packet and the dashed line is for the external potential. Figure~\ref{f6} shows the corresponding momentum space profile of the wave packet starting from the ground-state one in the external potential $V_{\mathrm{ext}}(x)=\frac{1}{2}x^2$. This snapshot was taken at $t=14$, which is sufficientry long for the wave packets to develop the interference pattern. Figure~\ref{f6} clearly shows the emergence of the notched structure from a more general initial condition than that in a realistic experimental setup.  

\section{Experimental Feasibility}
In this section, we make some remarks on the possibility of real
experiments with the BEC of neutral atoms. One of the striking features of the BEC experiments 
is the visibility of the momentum space. One can extract the momentum
 distribution of cold atoms by the time-of-flight (TOF) method\cite{Davis}. In the TOF method, while atoms with large momenta can travel far away, atoms with small momenta can travel only small distances from the origin. Therefore, the momentum distribution is detected directly and we can observe the small radiation ripple through the formation of the interference pattern with the aid of the TOF method. Moreover,
the control of external environments is relatively easy
in the BEC systems where we can confine condensate particles along a quasi-rectilinear line by tightening the laser beam trap. Therefore, the confirmation of the emitted radiation in BEC experiments is promising. As general requirements for this experiment, we note that a sufficiently large amplitude is required to form the 2-soliton bound state and the strength of the external potential is important. If one uses a very loose potential, the ground-state wave packet might have an insufficient amplitude and might fail to form the 2-soliton bound state. We recommend that the coefficient in front of $x^2$ be chosen as larger than $\frac{1}{2}$.
 
In addition, we must take special care for some subtleties arising from one-dimensional geometry and attractive interaction. In a recent study, Tovbis and Hoefer\cite{Tovbis} also treated the self-interaction and pattern formation of solitons in a similar environment and discussed the possible experimental condition. Although their argument is for the real space, which is different from our argument in the momentum space, we should also make considerations on the experimental feasibility. For that purpose, we rewrite the NLSE (\ref{NLSE}) back in the original form with dimensional constants as
\begin{equation}
\mathrm{i}\hbar\frac{\partial \psi}{\partial t}=-\frac{\hbar^2}{2m}\frac{\partial^2 \psi}{\partial x^2}-g_{1D}|\psi|^2\psi\label{dimension},
\end{equation}
where $\hbar$ is the Dirac constant, $m$ is the mass of the Bose-condensed atoms, and $g_{1D}$ is the coupling constant. In the typical three-dimensional geometry, $g_{3D}$ is written as $g_{3D}=\frac{4\pi\hbar^2 |a_{3D}|}{m}$, where $a_{3D}<0$ is the s-wave scattering length of the atoms with an attractive interaction. The s-wave scattering length is a quantity peculiar to a three-dimensional geometry and it is not a trivial problem to seek for the one-dimensional correspondent. However, in real experiments, we can stay in the NLSE regime provided that the radial confinement diameter $a_{\perp}$ is experimentally suitable level\cite{Strecker}.

 The next issue is related to the realization of eq.~(\ref{NLSE}). In a typical experiment, we initially trap atoms by applying a harmonic external field and suddenly switch off the trap and release atoms to recover the flat geometry along the axial direction at $t=0$. Strictly speaking, when we use an optical trap, a relatively loose Gaussian potential remains along the axial direction even after the release. However, its curvature can be neglected for our purpose. The harmonic external field defines the typical length scale of the system, $a_{\mathrm{ho}}=\sqrt{{\frac{\hbar}{m\omega}}}$, where $\omega$ is the trap frequency. We can reduce eq.~(\ref{dimension}) to eq.~(\ref{NLSE}) by replacing $x$ with $\sqrt{\frac{\hbar}{m\omega}}x$, $t$ with $\frac{t}{\omega}$, and $\psi$ with $\sqrt{\frac{N}{4\sqrt{\pi}}}\psi$, where $\mathit{N}$ is the total number of particles, and by letting $\frac{Ng_{1D}}{4\sqrt{\pi}\hbar\omega}$ be unity. It follows that 
\begin{equation}
N=\frac{4\sqrt{\pi}\hbar\omega}{g_{1D}}\label{N1}.
\end{equation}
More generally, if we let $\frac{Ng_{1D}}{4\sqrt{\pi}\hbar\omega}$ be $g'_{1D}$, a dimensionless coupling constant, eq.~(\ref{NLSE}) is generalized to be
\begin{equation}
\mathrm{i}\psi_{t}=-\frac{1}{2}\psi_{xx}-g'_{1D}|\psi|^2\psi,
\end{equation}
and we obtain
\begin{equation}
N=\frac{4\sqrt{\pi}\hbar\omega g'_{1D}}{g_{1D}}\label{N2}.
\end{equation}
In real experiments, it would be difficult to adjust $g'_{1D}$ exactly to unity. For such cases, we give here the tolerance for $g'_{1D}$ to detect the desired fringe. Figure~\ref{f7} shows the profile of the ground-state wave packet in the external potential $V_{\mathrm{ext}}(x)=\frac{1}{2}x^2$ with $g'_{1D}$ kept at 0.7, and Fig.~\ref{f8} shows the corresponding momentum space profile of the wave packet starting from the ground-state one in the external potential. This is a snapshot taken at $t=14$. It is found that the contrast of the fringe is relatively low because a smaller coupling constant $g'_{1D}$ makes the role of nonlinear oscillation less important. Therefore, we can roughly estimate that the lower bound of the suitable $g'_{1D}$ is about unity.

Next is the upper bound. Figure~\ref{f9} shows the profile of the ground-state wave packet in the external potential $V_{\mathrm{ext}}(x)=\frac{1}{2}x^2$ with $g'_{1D}$ kept at 5, and Fig.~\ref{f10} shows the corresponding momentum space profile of the wave packet starting from the ground-state one in the external potential. This also shows a snapshot taken at $t=14$. In this case, a strong self-focusing interaction makes the initial profile narrower and this leads to a wide-spread profile in the momentum space. In this case, the image intensity in the momentum space would be lower. We should keep image intensity above the detectable level, and it defines the upper bound of $g'_{1D}$. As a whole, the suitable range for the fringe-detection experiment is roughly estimated to be about $1<g'_{1D}<5$.

Finally, let us consider preparing initial condition (\ref{init2}).
As previously mentioned, we can change $g_{1D}$ freely by the Feshbach resonance technique to gain a sufficient number of atoms, as seen from eqs.~(\ref{N1}) and (\ref{N2}). However, the order of $N$ is upper-bounded, because this quantity must satisfy the inequality
\begin{equation}
 N<0.67\frac{a_{\perp}}{|a_{3D}|},
\end{equation}
to avoid the self-collapse of the condensate owing to the attractive interaction\cite{Parker}. In real experiments, Strecker {\it et al.} used the values $a_{\perp}\approx 1.5~\mathrm{\mu m}$ and $a_{3D}\approx -0.16~\mathrm{nm}$, letting the order of $N$ be several thousand. Therefore, the experimental feasibility of our proposal seems to be promising with the aid of the Feshbach resonance technique, provided that the number of particles is kept moderate. 

\section{Discussion and Summary}
As evidence that supports our simplified model, we can consider that our model explains why ``the higher the wave number, the narrower the pitches of the interference pattern,'' as shown in Fig.~\ref{f10}. This is due to the term including $\mathrm{cos}\frac{1}{2}k^2t$ in eq.~(\ref{result}), which validates the low-amplitude-defusing wave approximation of eq.~(\ref{LAD}). 

Next, the oscillations in the momentum space itself are not rare phenomena. As an example, we can immediately give the rectangular function which leads to the sinc function after the Fourier transformation. However, such functions tend to be angular, in general, and to exhibit oscillations in the momentum space. On the contrary, smooth functions, such as the Gaussian and sech-type ones, do not exhibit oscillations even after the Fourier transformation. Our wave packets which smoothly and monotonically decrease from the origin are among them and are not angular. Therefore, ``interference'' by our model is the most promising mechanism to explain the oscillation in the momentum space.

We have proposed an indirect method for observing radiation from an incomplete soliton.
According to this method, radiation forms the notched structure on the surface
profile of the wave packet in the momentum space with the aid of the TOF method. The origin of this structure
is the result of the interference between the main body of the oscillating soliton and the small radiation
in the momentum space. We numerically integrated the NLSE and Fourier-transformed it to confirm
 that the predicted structure really appears. Our concept was successfully proved and our simple
 model reproduced the qualitative result from the exact numerical simulation.  
Finally, we discussed the possible conditions required in real experiments with BEC.

\section*{Acknowledgment}
One of the authors, H.~F. thanks Utsunomiya University for the hospitality and for offering wonderful working spaces and
opportunities of fruitful discussion.

\appendix
\section{}
In this work, we consider initial conditions that slightly deviate from that of the exact 2-soliton bound state (\ref{init1}). Their eigenvalue spectra should include 2 discrete eigenvalues. The typical example is the Gaussian wave packet (\ref{init2}). In fact, initial conditions (\ref{init1}) and (\ref{init2}) appear very similar, as shown in Fig.~\ref{f1}. The stationary solutions formed in the harmonic potentials are also considered to meet the above conditions.

Initial condition (\ref{init1}) evolves into eq.~(\ref{2sol}). The purpose of this appendix is to explain the validity of the assumption of superposition we made in eq.~(\ref{super}). Under this assumption, general wave packets that meet the above conditions can be approximated as the sum of the oscillating main part and the small diffusing part. The smallness of the latter enables us to express it as a general solution of the linear Schr\"odinger equation.
 
Consider the linear potential scattering problem (\ref{eigen-1}) and give the corresponding Jost functions as 
\begin{displaymath}
\phi(x;\xi)=\left(\begin{array}{cc} 1 \\ 0 \\ \end{array} \right)\mathrm{exp}(-i\xi x),	x \to -\infty.
\end{displaymath}
\begin{displaymath}
\psi(x;\xi)=\left(\begin{array}{cc} 0 \\ 1 \\ \end{array} \right)\mathrm{exp}(i\xi x),	x \to \infty.
\end{displaymath}
\begin{displaymath}
\bar{\psi}(x;\xi)=\left(\begin{array}{cc} 1 \\ 0 \\ \end{array} \right)\mathrm{exp}(-i\xi x),	x \to \infty,
\end{displaymath}
where $\xi$ is the eigenvalue of the linear scattering problem (\ref{eigen-1}) and consists of discrete and continuous parts. We write $\phi(x;\xi)$ as a linear combination of $\psi(x;\xi)$ and $\bar{\psi}(x;\xi)$,
\begin{displaymath}
\phi(x;\xi)=a(\xi)\bar{\psi}(x;\xi)+b(\xi)\psi(x;\xi),
\end{displaymath}
where the transmission and reflection coefficients $a(\xi)$ and $b(\xi)$ satisfy 
\begin{displaymath}
|a(\xi)|^2+|b(\xi)|^2=1.
\end{displaymath}
From the general arguments of the potential scattering theory, $\xi$ can be analytically continued to the upper complex plane, and bound states correspond to the discrete zero points of $a(\xi)$,
\begin{displaymath}
a(\zeta_{n})=0.
\end{displaymath}
According to Zakharov and Shabat\cite{Zakharov}, the solution $u(x,t)$ of the original NLSE (\ref{NLSE}) can be written as
\begin{displaymath}
u(x,t)=2\sum_{n}\left(\sqrt{\frac{b(\zeta_{n})}{a'(\zeta_{n})}}\right)^{\ast}\mathrm{exp}(-i\zeta_{n}x+i\zeta_{n}^{2}t)\psi_{n2}^{\ast}+\frac{1}{i\pi}\int_{-\infty}^{\infty}\Phi_{2}^{\ast}(\xi)d\xi.
\end{displaymath}
Here, the zeros of $a(\xi)$ are assumed to be simple poles and $a'(\xi_{n})=\frac{\partial a}{\partial \xi}|_{\xi=\xi_{n}}$ is not equal to zero. In the above formula, the discrete part including $\psi_{n2}$ corresponds to the soliton part ($n=1,2$ for the 2-soliton bound state) and the integral part including $\Phi_{2}$ to the small diffusing wave. This is clearly a superposed form of the soliton-like and  small-diffusing-wave parts. Neither can be expressed analytically. The integrand $\Phi_{2}(\xi)$ is the quantity that is determined by the simultaneous equations below.
\begin{displaymath}
\Phi_{1}-c(x,\xi)\left[\frac{1}{2}(1+\hat{T})\right]\Phi_{2}^{\ast}=-c(x,\xi)\sum_{n}\frac{\lambda_{n}^{\ast}}{\xi-\zeta_{n}^{\ast}}\psi_{n2}^{\ast},
\end{displaymath}
\begin{displaymath}
c^{\ast}(x,\xi)\left[\frac{1}{2}(1-\hat{T})\right]\Phi_{1}+\Phi_{2}^{\ast}=c^{\ast}(x,\xi)\left(1+\sum_{n}\frac{\lambda_{n}}{\xi-\zeta_{n}}\psi_{n1}\right),
\end{displaymath}
\begin{displaymath}
\frac{\psi_{m1}}{\lambda_{m}}+\sum_{n}\frac{\lambda_{n}^{\ast}}{\zeta_{m}-\zeta_{n}^{\ast}}\psi_{n2}^{\ast}=\frac{1}{2\pi i}\int_{-\infty}^{\infty}\frac{\Phi_{2}^{\ast}}{\xi-\zeta_{m}}\xi,
\end{displaymath}
\begin{displaymath}
-\sum_{n}\frac{\lambda_{n}}{\zeta_{m}^{\ast}-\zeta_{n}}\psi_{n1}+\frac{\psi_{m2}^{\ast}}{\lambda_{m}^{\ast}}=1+\frac{1}{2\pi i}\int_{-\infty}^{\infty}\frac{\Phi_{1}}{\xi-\zeta_{m}^{\ast}}\xi,
\end{displaymath}
where we introduced the abbreviations
\begin{displaymath}
\lambda_{n}=\sqrt{\frac{b(\zeta_{n})}{a'(\zeta_{n})}}\mathrm{exp}(i\zeta_{n}x-i\zeta_{n}^{2}t),
\end{displaymath}
and
\begin{displaymath}
c(x,\xi)=\frac{b(\xi)}{a(\xi)}\mathrm{exp}(2i\xi x-2i\xi^{2}t).
\end{displaymath}
The Hilbert transformation operator $\hat{T}$ acts as
\begin{displaymath}
\hat{T}\Phi=\frac{1}{i\pi}\int_{-\infty}^{\infty}\frac{\Phi(\xi')}{\xi'-\xi}d\xi.
\end{displaymath}
Since the pure soliton-like eq.~(\ref{init1}) acts as the reflectionless potential in eq.~(\ref{eigen-1}), $b(\xi)$ should be zero. This makes $c(x,\xi)=0$, and we can easily see $\Phi_{1}=\Phi_{2}=0$ from the above simultaneous equations. Then, the small-diffusing-wave part completely disappears and only the discrete soliton part corresponding to eq.~(\ref{2sol}) remains. This is a pulsar localized around the origin and oscillates with period $\frac{\pi}{4}$.

In this paper, we consider the initial conditions that slightly deviate from that of the exact 2-soliton bound state (\ref{init1}). From the principle of continuity, they are considered to evolve very similarly to pulsars (exact 2-soliton bound states). Because the variation in magnitude at the origin is from 2 to 4, we have modeled this pulsar as eq.~(\ref{app}). This function is localized at the origin and oscillates there with period $\frac{\pi}{4}$. It satisfies the condition to be a model function of the exact 2-soliton bound state (\ref{init1}). By numerical integration, we have already confirmed that the wave packets starting from the initial conditions that slightly deviate from that of the exact 2-soliton bound state well retain the properties of a solitary wave\cite{Fujishima}. As a soliton-like part, we can use eq.~(\ref{2sol}) instead of eq.~(\ref{app}). However, this function cannot be Fourier-transformed analytically and can be unsuitable as a brief model.

The amplitude due to the integral part is considered to be small because the integrand $\Phi_{2}$ is small as long as $b(\xi)$ and $c(\xi)$ are small. Therefore, this part can be well approximated by the general solution of the linear Schr\"odinger equation (\ref{LAD}).

\end{document}